# Sub-40mV Sigma-VTH IGZO nFETs in 300mm Fab

Jerome Mitard, Luka Kljucar, Nouredine Rassoul, Harold Dekkers, Michiel van Setten, Adrian Vaisman Chasin, Geoffrey Pourtois, Attilio Belmonte, Gabriele Luca Donadio, Ludovic Goux, Ming Mao, Harinarayanan Puliyalil, Lieve Teugels, Diana Tsvetanova, Manoj Nag, Soeren Steudel, Jose Ignacio del Agua Borniquel[2a], Jothilingam Ramalingam[2], Romain Delhougne, Chris J.Wilson, Zsolt Tokei, Gouri Sankar Kar

imec, Kapeldreef 75, B – 3001 Leuven, Belgium
[2] Applied Materials, MPD, Santa Clara, USA
[2a] also imec resident, Kapeldreef 75, B – 3001 Leuven, Belgium

Back and double gate IGZO nFETs have been demonstrated down to 120nm and 70nm respectively leveraging 300mm fab processing. While the passivation of oxygen vacancies in IGZO is challenging with an integration of front side gate, a scaled back gated flow has been optimized by multiplying design of experiments around contacts and material engineering. We then successfully demonstrated sub-40mV σ($V_{TH\_ON}$) in scaled IGZO nFETs. Regarding the performance and the $V_{TH\_ON}$ control, a new IGZO phase is also reported. A model of dopants location is proposed to better explain the experimental results reported in literature.

## Introduction

In today's highly information-oriented society, groundbreaking hardware innovations targeting Augmented Intelligence are necessary to compute on even-more limited power budget. This requires higher density memory as well as smart interconnect solutions that can actively reconfigure leveraging the 3D dimension [1]. $InGaZnO_4$ (IGZO) as a thin film channel material for FETs can become a key element for the aforementioned applications since its premise relies on 1) zepto ampere off-state leakage capability namely few electrons flowing per FET element in one year [2], 2) relatively good electron mobility especially when compared to that of doped-amorphous Si [3] and 3) low thermal budget processing to possibly enable a sequential integration with conventional Si-based transistors. In this work, IGZO device integration is reported leveraging our 300mm-fab facilities. Our objective is mainly to gain insights into the process and material elements which drive the control of the performance parameters of IGZO nFETs.

## Fabrication Of Scaled 300mm-IGZO NFET

Process details are given in Figure 1. IGZO nFET consists of three main modules: active area, source/drain contacts and front side gate in case of double gated nFETs. First, a highly doped Si 300mm wafer with an option of blanket metal deposition serves as a common back-gate for the transistors. A thick (SiCN/) ALD $Al_2O_3$ is used as bottom gate dielectric. The next step utilizes 300mm AMAT Endura® Impulse PVD IGZO. Active patterning is applied on the entire stack down to the Si substrate. The gate is formed by patterning

TiN/W stack down to Al$_2$O$_3$ gate dielectric level. This minimizes the damages induced by gate patterning. The last module is the Source and Drain contact formation through a PECVD SiO$_2$ field oxide. Ti-based/W metallization is applied similarly to [4].

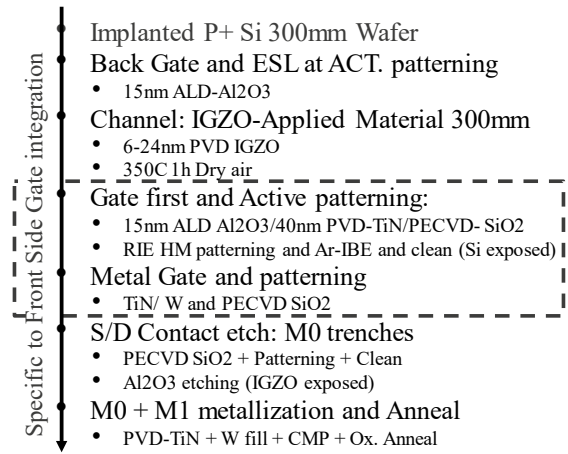

Figure 1. Detailed flow description of back and double IGZO nFETs.

It is worthwhile mentioning that devices have been successfully demonstrated up to M1 level enabling the study of small device dimensions. Figure 2 Left shows fully functional 70nm-Lg double gate and 120nm-Lg back gate IGZO nFETs. Corresponding xTEM images are given in Figure 2 -Right panel as reference.

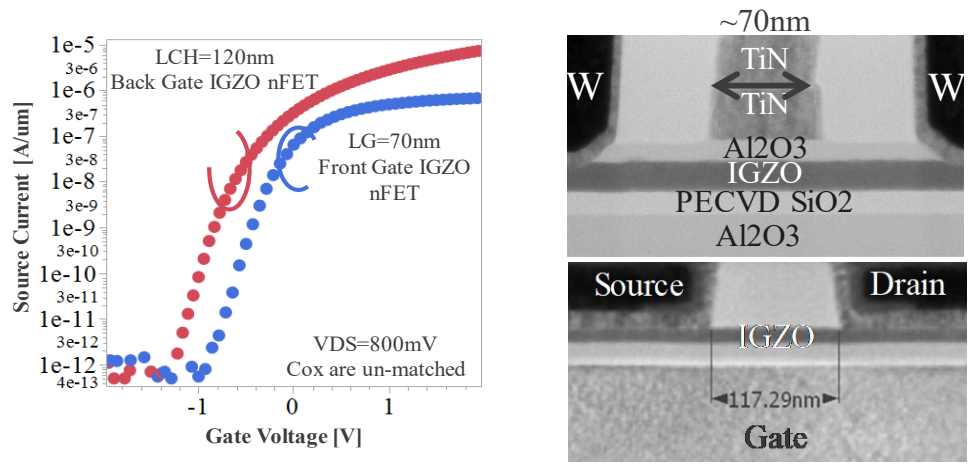

Figure 2. Left. IVs of 70nm-front side and 120nm back-gate IGZO nFETs. Right-top: xTEM of 70nm-front side IGZO nFETs. Bottom-Right: xTEM of 120nm-Back gate IGZO.

# Discussion About Variability In IGZO NFET

N-type doping depends on the IGZO crystallographic structure

There are two main electron doping mechanisms known in amorphous-IGZO: a sub stoichiometric amount of oxygen [oxygen vacancies (Vo)] and the incorporation of hydrogen. Both give very shallow donor levels in the bandgap and then act as n-type dopants in IGZO. The weakest bonded oxygen atoms like those that are under-coordinated or that are involved in multiple Zn-O bonds [5] can be released in presence of hydrogen to form OH ions [6]. Fig. 3 shows the impact of a mild 10% H anneal at >200°C.

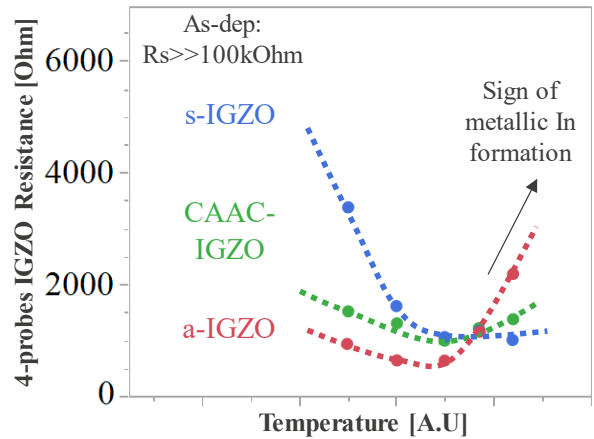

Figure 3. Resistivity measurements after 1h FGA (10% $H_2$). 50nm IGZO different phases and temperatures.

A clear drop of the IGZO resistivity with respect to the control samples (as-deposited) is in line with a doping increase. More interestingly, we found that the initial response to hydrogen anneal can be modulated by the crystallographic structure of IGZO.

The challenge of passivating oxygen vacancies w. top gate

To control the final doping of the IGZO channel, it is well reported in literature [7] that a post processing oxygen anneal can be applied to passivate the oxygen vacancies which are formed during the device fabrication. This technique is also used in our 300mm flow as described in Fig. 1. As it might be expected, in front gate IGZO nFETs, the passivation efficiency is limited by the presence of the top gate stack. Therefore, it seems important to limit the formation of the oxygen vacancies during the deposition of the gate dielectrics. Three oxides were studied: $SiO_2$, $Al_2O_3$ and $HfO_2$. The experiment consists of measuring the IGZO resistance before and after oxide deposition. Figure 4 confirms that a low resistivity IGZO is measured after oxide deposition. After oxygen annealing, only PECVD-$SiO_2$ shows a large recovery while for $Al_2O_3$ and $HfO_2$, it remains low.

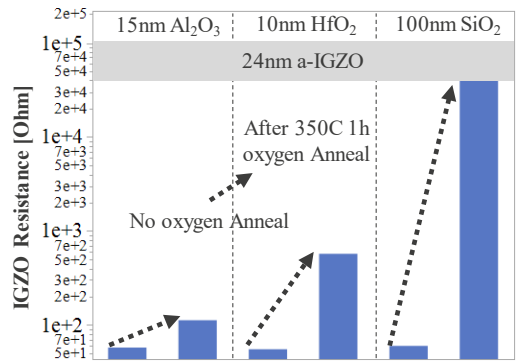

Figure 4. 4p-Resistance measurement of CTLM devices (10um-spacing). Effect of the oxygen Anneal in presence of IGZO capping.

This result clearly challenges the implementation of conventional water-based high-K materials in gate first IGZO integration. However, a high $I_{ON}/I_{OFF}$ ratio can still be obtained with highly doped (i.e. un-passivated Vo) IGZO nFETs by increasing the gate coupling. Fig. 5 shows an experimental demonstration of a 6nm-thick doped-IGZO nFETs down to 120nm-Lg. Since ultra-thin IGZO integration has its own challenge in term of variability and process controllability, we primarily focus on back gated IGZO nFET from now on.

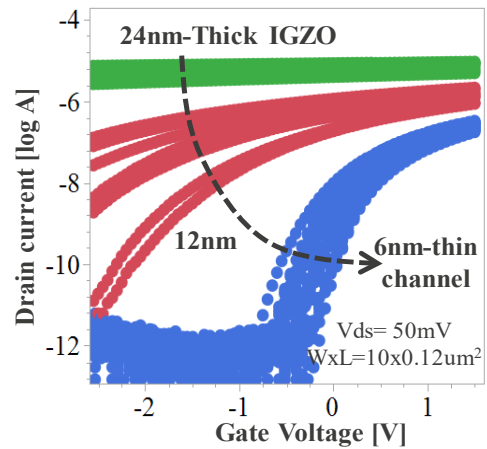

Figure 5. Effect of IGZO thickness on electrostatic control. Full carrier depletion can be obtained.

Contact engineering for IGZO nFET with Ti metal barrier

Contrary to the use of undoped IGZO in the channel for $I_{OFF}$ control, contacts can rely on maximizing the oxygen vacancies to increase the dopant concentration in the S/D regions. When not done locally, it could be an extra source of variability. This doping is made through oxygen scavenging from IGZO by a thin Ti layer between the liner (TiN) and the IGZO. Low specific contact resistivity down to $1 \times 10^7$ Ohm.cm$^2$ is demonstrated when the Ti layer is thinner than 5nm (Fig. 6). With thicker barrier, the formation of $TiO_2$ and specific alloys is taking place at the IGZO/Contact interface. This is both confirmed by ab initio simulations (Fig. 7) and by advanced physical characterization techniques (see Fig. 8).

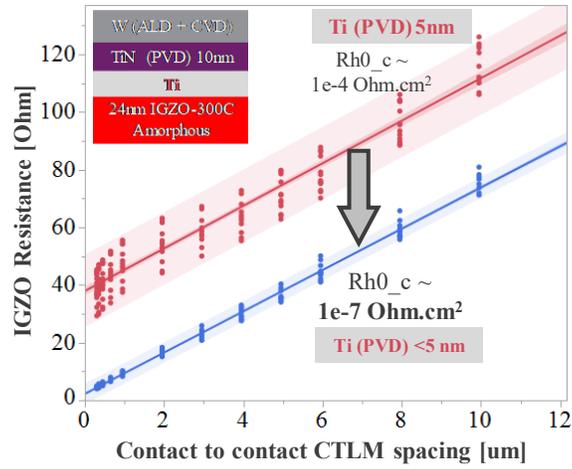

Figure 6. IGZO contact engineering with Ti thickness optimization.

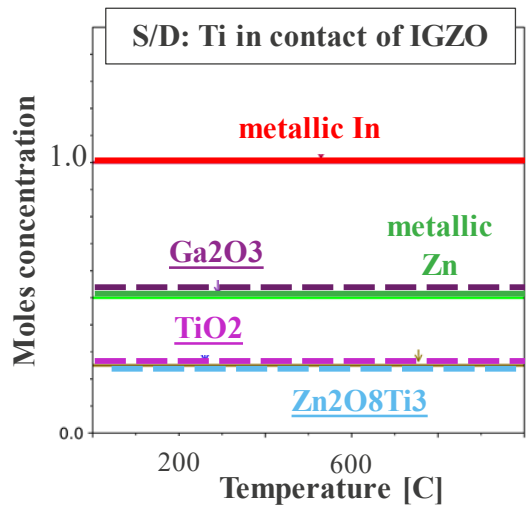

Figure 7. Ab-Initio (DFT@PBEsol) modeling of the IGZO/Ti interface.

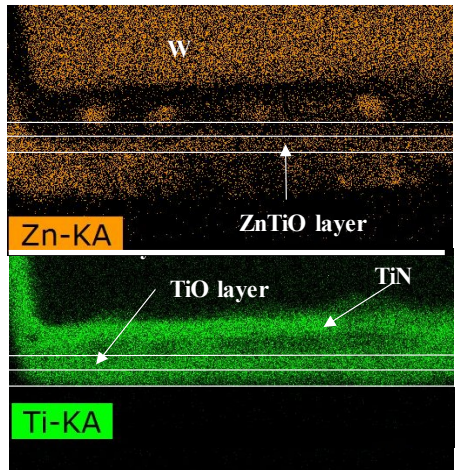

Figure 8. EDS characterization of the S/D contact region when 10nm-thick Ti is in contact with IGZO-PVD.

Amorphous IGZO, C-Axis Aligned IGZO and new s-IGZO

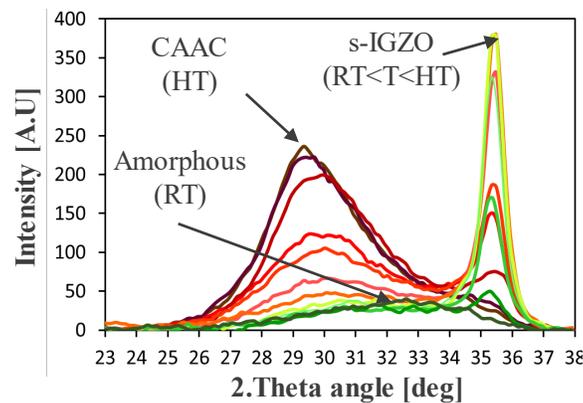

Figure 9. XRD characterization of blanket IGZO when chuck temperature is varied from Room Temperature to High Temperature.

X-ray diffraction (XRD) techniques are used extensively since this is a nondestructive in-fab technique that provides detailed information about the crystallographic structure of the IGZO material. Figure 9 shows a typical spectrum where clear peaks/humps are seen and attributed to: 1. amorphous IGZO, 2. CAAC-IGZO and 3. to the best of our knowledge, previously not reported new phase called here s-IGZO (spinel phase). The s-phase is only formed under certain conditions of power, temperature, and oxygen flow during material deposition. In parallel, we verified that the transition between the different phases is not due to any compositional changes within the IGZO material.

Thick IGZO (>12nm) backgated nFETs with active layers submitted to final $O_2$ anneal are used to study the different phases of IGZO and their electrical impact on device parameters. In this configuration, the carrier transport preferentially occurs in the bottom half of the IGZO channel while the top half ($SiO_2$/IGZO interface) mostly drives the electrostatic control of BG transistors. In Fig.10, $I_{ON}$-$V_{TH\_ON}$ long channel trends are shown and highlight a clear difference between CAAC- and a-IGZO already.

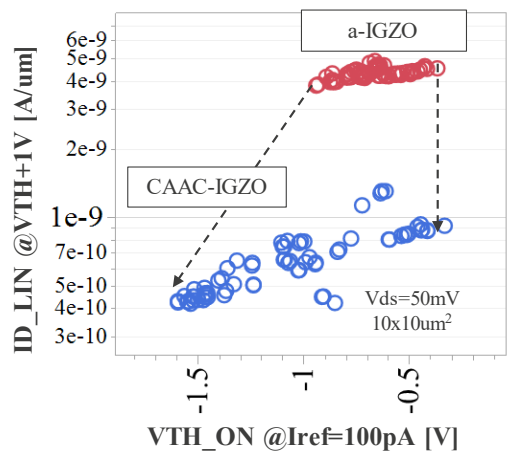

Figure 10. Degradation of ID in linear regime, taken at offset $V_{TH\_ON}$ and increased variability for 24nm-CAAC-IGZO.

The amorphous IGZO has much reduced spread in VTH-ON and higher ID,LIN (~mobility) than CAAC-IGZO. Reliability tests have also been carried out to compare the two phases and the results show the existence of two competing PBTI degradation mechanisms for CAAC-IGZO (Figure 11). Notice that it is important to report the degradation over time and different gate voltage stress because a cut line at a specific value could have shown no BTI degradation of CAAC-IGZO and then attributing an unfair benefit of this phase over the a-IGZO phase. Combining these findings to the result shown in Fig. 3 where the sheet resistance of IGZO under hydrogen exposure depends on the phases, we can reasonably conclude that CAAC owns different doping levels in comparison to a-IGZO.

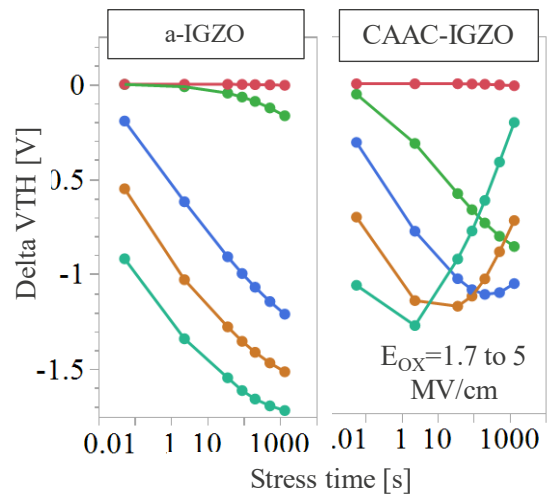

Figure 11. Positively-charged trapping only observed in CAAC-IGZO ("Si-like" PBTI).

To gain insight into longitudinal doping (from Source to Drain), a short channel study is performed. The impact of an increased oxygen partial pressure during the PVD deposition is first reported for both a-IGZO and CAAC-IGZO. Figure 12 - left side - demonstrates that a constant increase of the oxygen partial pressure during amorphous-IGZO deposition results in a continuous increase of $I_{ON}$ and $V_{TH\_ON}$. This is in line with a reduction of oxygen vacancies and therefore a lowering of effective n-type doping in the thin film [7]. On the contrary, CAAC-IGZO formed at temperature larger than 25°C and for a positive $O_2$% flow show reversed trends namely a reduced $I_{ON}$ as well as $V_{TH\_ON}$ from low to high oxygen content (Fig. 12 -Right panel-).

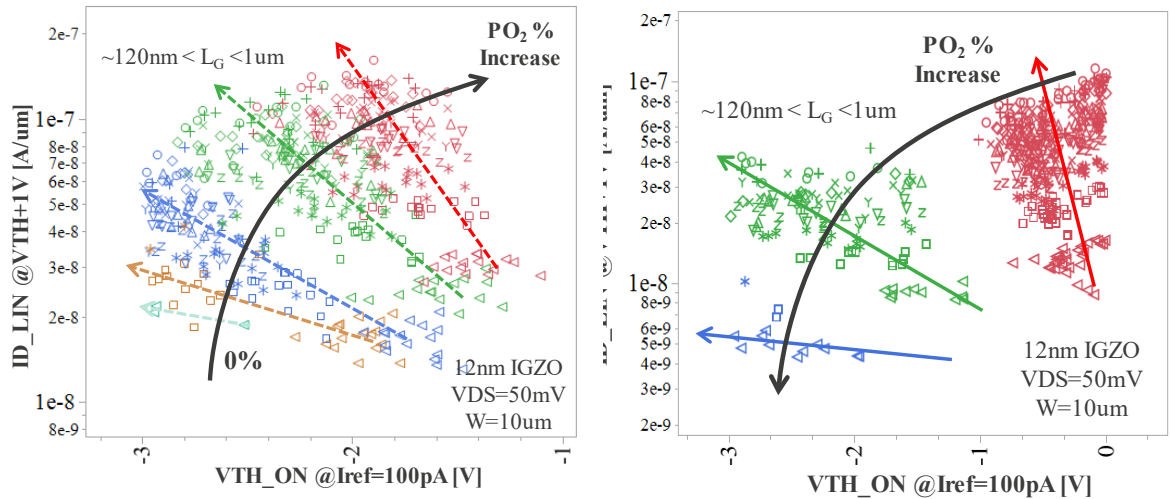

Figure 12. (Left) Increased flow of oxygen deposition improves VTH_ON and Idlin performance of a-IGZO. (Right) Increased flow of oxygen during deposition degrades VTH_ON and Idlin performance of CAAC-IGZO

In CAAC-IGZO, supplying oxygen in excess leads to a severe degradation of the performance of the transistors, especially when compared to Room Temperature (RT) amorphous IGZO. The temperature during IGZO deposition can also be tuned to obtain s-IGZO. Fig.13 shows that short channel effect control of s-IGZO nFETs have the best trade-off between $I_{ON}$ (mobility) and $V_{TH\_ON}$ (doping) over the two other IGZO structures (CAAC and a-IGZO).

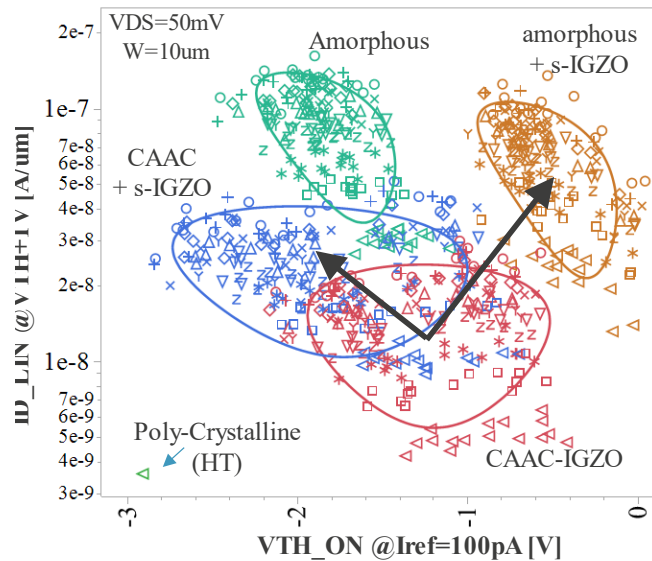

Figure 13. The combo of 12nm- aIGZO and the new s-IGZO gives the best trade-off VTH_ON short channel performance.

Increased mobility in s-IGZO has been studied by ab-initio simulations and we confirm that this specific crystalline structure leads to free carriers with lower effective mass and consequently higher mobility than with CAAC-IGZO (see Fig. 14).

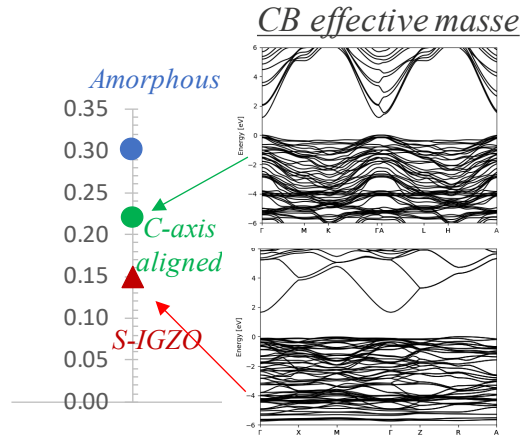

Figure 14. Ab-initio simulations confirming the s-phase of IGZO has reduced effective mass as compared to CAAC. The effective mass of the two crystalline phase is obtained from the calculated band-structure, the effective mass for a-IGZO is taken from literature [8].

Since the back gated architecture allows a profiling of dopant within IGZO, a sequential deposition of a-IGZO and semi-crystalline IGZO has been attempted to see if both the carrier transport and the short channel effects control can be improved. A 6nm amorphous layer followed by a 6nm low O%-CAAC IGZO should bring the short channel IGZO nFETs to the high-performance side. However, depositing the CAAC IGZO layer on the amorphous phase IGZO causes severe device degradation for all different layer thickness ratios (see Fig.15). This result is explained by the thermal budget used to create the CAAC structure which is incompatible with RT a-IGZO. One-step s-IGZO thin-film deposition seems an attractive option for obtaining high performance short channel IGZO nFETs.

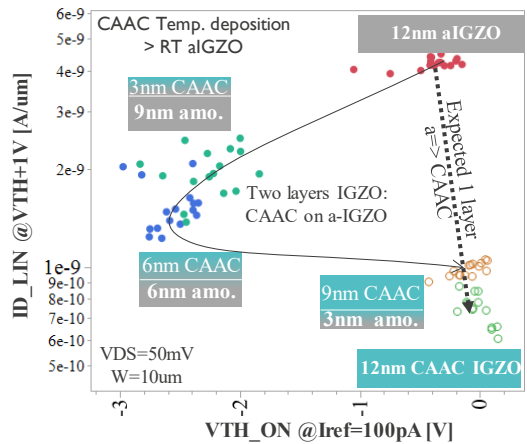

Figure 15. Dual IGZO deposition CAAC on top of a-IGZO. Higher thermal budget of the second layer degrades the bottom a-IGZO.

The lateral dopant profiling is done through the study of the effect of narrowing the channel width of IGZO. 20nm-thick low O% CAAC IGZO nFET is selected because we revealed moderate $V_{TH\_ON}$ (LCH) roll-off, meaning a reduced amount of Vo in the top half of the IGZO channel if compared to a-IGZO. Fig. 16 shows the effect of WCH scaling on $V_{TH\_ON}$ where a drastic improvement of electrostatic control is observed for the transistor width smaller than 200nm. This result highlights a strong non-uniformity of n-type dopant concentration along the width of the device.

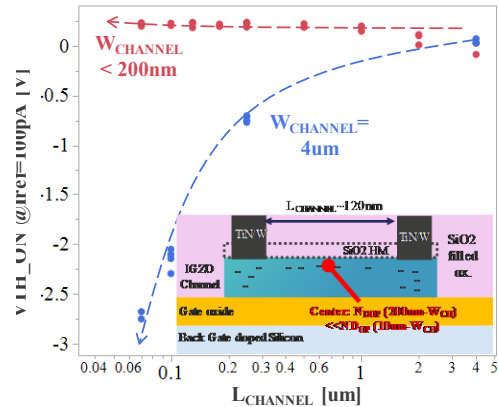

Figure 16. Narrowing the IGZO channel drastically increases the controllability of $V_{TH\_ON}$.

A Scanning Spreading Resistance Microscopy (SSRM) analysis has been performed on test samples (Fig. 17) and a reduced resistivity is observed near the edge of the IGZO active area. Note that the effect cannot be seen along $L_{CHANNEL}$ due the high n-type doping created by the S/D metallization. An empirical model has been built to gather all the information from the previous experiment (Fig. 17).

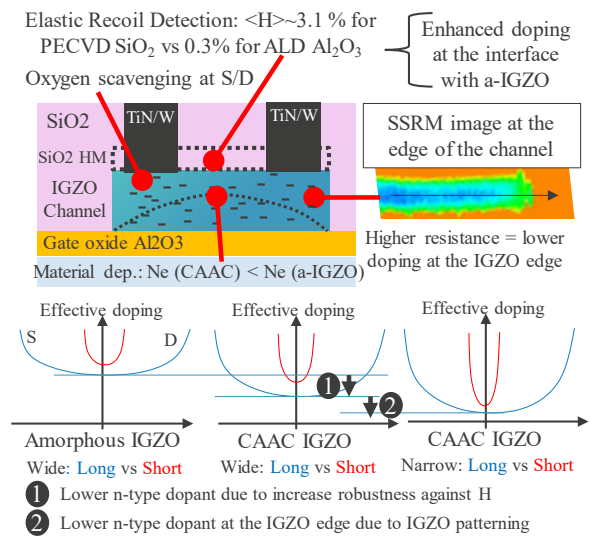

Figure 17. Dual IGZO deposition CAAC on top of a-IGZO. Higher thermal budget of the second layer degrades the bottom a-IGZO.

## Demonstration Of σ(VTH-ON) Down To 20mV

Based on the previous learning about n-type dopant location in IGZO, we show that the performance of transistors keeps increasing when the channel length is reduced and scales with the channel width (Fig. 18). A measure of the $V_{TH-ON}$ variation has been performed at the 300mm wafer scale from long to ~120nm $L_{CH}$ and down to 200nm $W_{CH}$ dimensions. Fig. 19 shows the Id-Vg curves of >100 Back Gated IGZO-nFETs with no failed devices detected. The standard variation of the $V_{TH-ON}$ across $L_{CH}$ and $W_{CH}$ is often less than 40mV with a minimum of 20mV (Fig. 20). The NBTI reliability has been evaluated on these Vo-controlled back gated IGZO nFETs. Limited $V_{TH-ON}$ degradation under NBTI stress has been detected up to 1000s stress time and at oxide field up to 5MV/cm (Fig. 21).

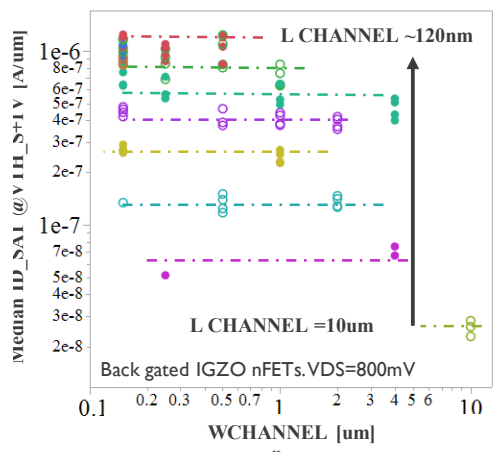

Figure 18. $I_{D-SAT}$ at offset $V_{TH-SAT}$ keeps increasing with $W_{CHANNEL}$ is scaled down.

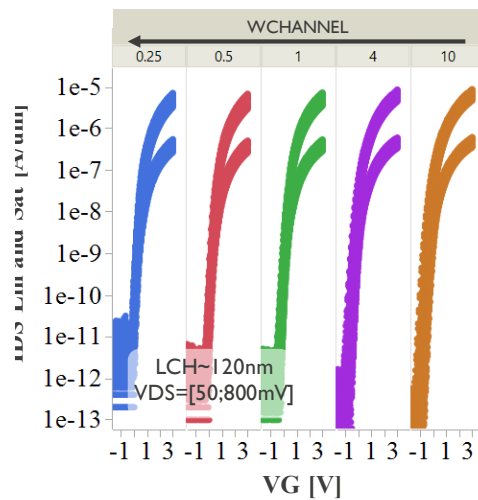

Figure 19. More than 100 back-gated IGZO nFETs functional across $W_{CH}$ dimensions. <u>No filtering process applied.</u>

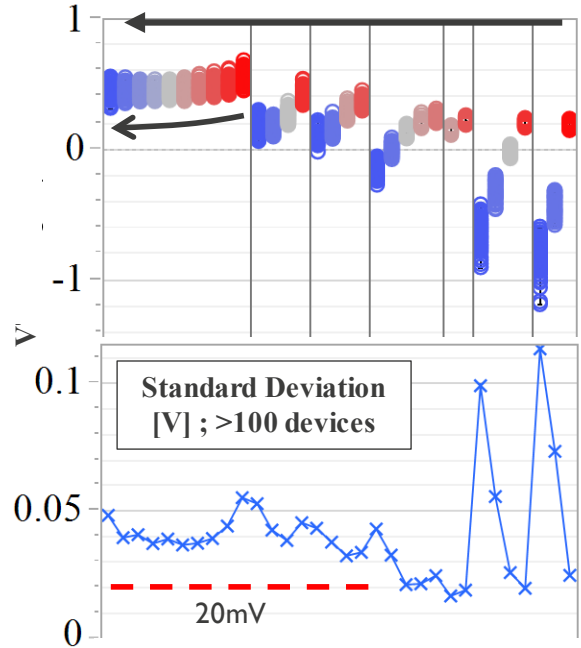

Figure 20. More than 100 back-gated IGZO nFETs functional across $W_{CH}$ dimensions. No filtering process applied.

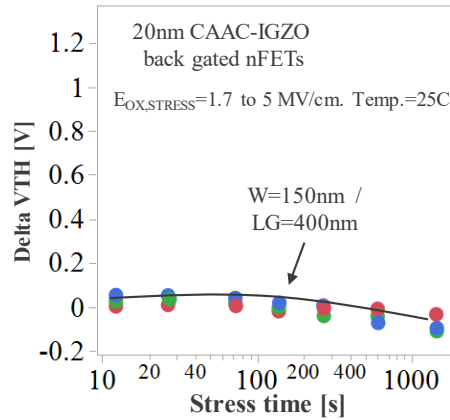

Figure 21. Stress and Sense NBTI degradation on narrow IGZO nFETs at $V_{TH\_ON}$ overdrive up to 11V.

## Conclusions

We have demonstrated scaled IGZO nFETs with excellent $V_{TH\_ON}$ control using an industry compatible 300mm process flow. This has been achieved thanks to a careful mapping of n-type doping in the three dimensions of the IGZO channel. While semi-crystalline IGZO seems to be more robust against hydrogen than that of amorphous-IGZO, a new IGZO phase is found to help boost $I_{ON}$ at short channel. The demonstration of back-gate IGZO nFETs with low variability and relatively high drive will benefit to the top-gate architecture development. This will provide new opportunities for the IGZO-based devices like an "on-the-fly" $V_{TH}$ setting for the performance control of advanced applications.

## Acknowledgements

The imec's Logic and Memory partners involved in the NanoIC and Emerging Memory programs, EU (project funded under the grant agreements: 687299 (NeuRam3-Horizon 2020) and 826655 (Tempo-Electronic Components and Systems for European Leadership Joint Undertaking), the pilot line and amsimec (test lab) are acknowledged for their support. A special thank you go to the numerous colleagues from the Unit Process and MCA teams as well as Y Cao & D. L.Diehl from Applied Materials Inc. Si Syst., MPD, Sunnyvale and A.Cockburn, AMAT-Belg. This work is a broad team effort.